%% file: main.tex
\documentclass[
    aps,
    prb,
    twocolumn,
    floatfix,nofootinbib,
    superscriptaddress
]{revtex4-2}

\usepackage{placeins}
\usepackage{multirow}
\usepackage{verbatim}

\usepackage[utf8]{inputenc}
\usepackage{times}
\usepackage[breaklinks]{hyperref}
\usepackage{color}
\usepackage{siunitx}
\usepackage{booktabs}
\usepackage[version=4]{mhchem}

\usepackage{amssymb} 
\usepackage{amsmath}
\usepackage{amsthm}
\usepackage{braket}

\input{commands}

\usepackage{tikz}
\usepackage{xcolor}
\usepackage{graphicx}

\usepackage[capitalise]{cleveref}

\makeatletter
\renewcommand\@biblabel[1]{#1.}
\makeatother

\begin{document}

\setcounter{secnumdepth}{2} 


\title{Chiral response of spin spiral-states as the origin of chiral transport fingerprints of spin textures}

\author{Jonathan Kipp}
    \affiliation{\pgi}
    \affiliation{\aachen}

\author{Fabian R. Lux}
    \affiliation{\mainz}
    
\author{Yuriy Mokrousov}

    \affiliation{\pgi}
    \affiliation{\mainz}


\begin{abstract}
The transport properties of non-trivial spin textures are coming under closer scrutiny as the amount of experimental data and theoretical simulations is increasing.
To extend the commonly accepted yet simplifying and approximate picture of transport effects taking place in systems with spatially varying magnetization, it is important to understand the transport properties of building blocks for spin textures $-$ the homochiral spin-spiral states.
In this work, by referring to phenomenological symmetry arguments based on the gradient expansion, and explicit calculations within the Kubo framework, we study the transport properties of various types of spin-spirals in a  two-dimensional model with strong spin-orbit interaction. In particular, we focus on the contributions to the magnetoconductivity, the planar Hall effect and the anomalous Hall effect, which are sensitive to the sense of chirality of the spiral states. We analyze the emergence, symmetry, and microscopic properties of the resulting chiral magnetoconductivity, chiral planar Hall effect, and chiral Hall effect in terms of spin-spiral propagation direction, cone angle, spiral pitch, and disorder strength. Our findings suggest that the presence of spin-spiral states in magnets can be readily detected in various types of magnetotransport setups.
Moreover, the sizable magnitude of chiral contributions to the conductivity of skyrmions estimated from homochiral spirals implies that chiral, as opposed to topological, magnetotransport can play a prominent role for the detection of non-trivial spin textures.
\end{abstract}

\maketitle

\date{\today}

\section{Introduction}

\begin{figure*}[t!]
	    \includegraphics[width=0.85\hsize]{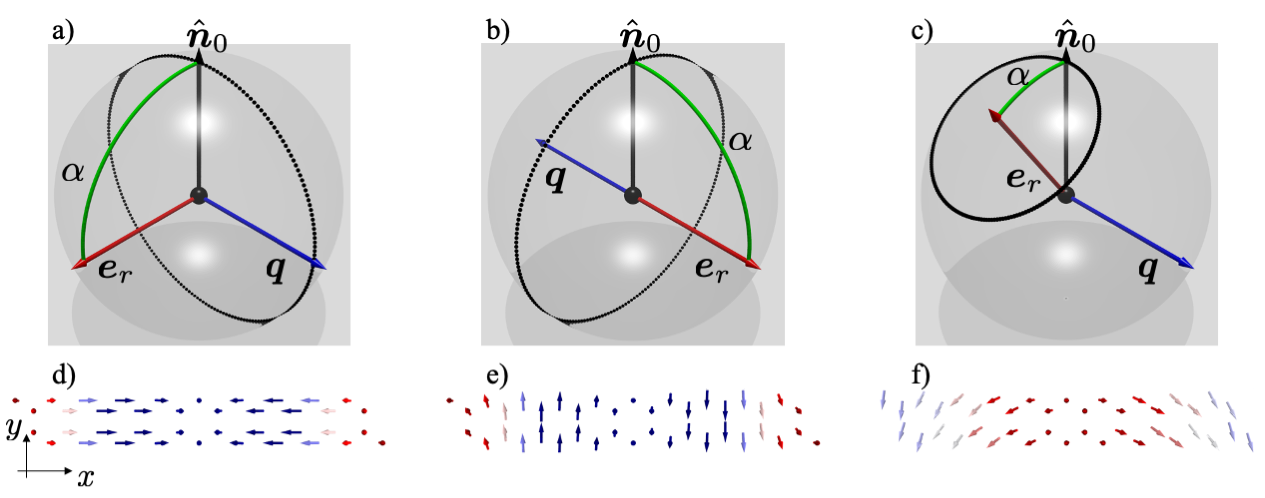}
		\caption{{\bf Parametrization of N\'eel, Bloch, and cone-type  spirals.}  Parts (a)-(c) show the parametrization of N\'eel (a), Bloch (b), and cone-type (c) spin-spirals on a sphere spanned by magnetic moments. The spiral propagates along a direction $\vec{q}$, and the magnetization rotates around an axis $\vec{e}_r$,  which encloses a cone angle $\alpha$ with the initial magnetic moment $\nuv{n}_0$. For the N\'eel  spiral $\vec{e}_r\bot\vec{q}$, while for the Bloch  spiral, $\vec{e}_r\parallel\vec{q}$. In this work, the plane of the honeycomb lattice is the $xy$-plane (axis $\nuv{z}$ is perpendicular to it) which contains $\vec{q}$.  Plots (d)-(f) show the real-space distribution (as seen from the top) of corresponding magnetization textures for $\vec{q}\parallel\nuv{x}$. }
		\label{Fig1}
\end{figure*}

The electron dynamics taking place in diverse magnetization textures is one of the most intensively pursued areas of solid state physics at the crossroads between topology, transport and physics of non-uniform media.
In particular, the transport manifestations of electron dynamics in spin textures exposed to external electric fields have come to occupy a very important role in modern skyrmionics and spintronics due to numerous implications for practical implementations of skyrmion-based design philosophy. 
Since the early days of the field, a commonly accepted approach for treating the transverse transport of complex spin-textures in two dimensions, such as skyrmions, has relied heavily on the presence of the so-called emergent field, coupling spatial gradients $\partial_x \hat{\mathbf{n}}$ and $\partial_y \hat{\mathbf{n}}$.
Since the seminal paper of Bruno and co-workers~\cite{Bruno2004}, the validity of the ansatz for the transverse Hall conductivity as an object directly proportional to the emergent field has been tested and validated in many cases from effective models~\cite{Tatara2002,Onoda2004,Nakazawa2018}, microscopic calculations~\cite{Franz2014,FreimuthSpencerPRB2018} and direct comparison to experiments~\cite{Neubauer2009,PfleidererRoschPRB2013}, while "topological" contributions have also been reported for magnetoconductivity and the planar Hall effect~\cite{Tokura2020PHE,FelserPRB2019,FelserPRB2020,FelserACS2021,GoebelReview2021}.
However, in recent years, evidence has started to accumulate that the simple picture of the emergent field and the corresponding topological Hall effect often does not suffice to explain results from experimental transport measurements. The community has become more aware of the fact that the emergent field picture has to be radically extended~\cite{Yamakage2021,TokuraPRB2021,JubaPRL2021}, particularly in strongly spin-orbit coupled systems.
On the one hand, recent works have shown that in spin-orbit coupled systems the second order in magnetization gradients terms, which go beyond the conventional topological-like signal, can contribute significantly to the orbital magnetism and Hall effect exhibited by the textures~\cite{Lux2018,JubaPRL2021}. 
On the other hand, it has been demonstrated that an effect that is much more pronounced
in strongly spin-orbit coupled systems, since it is linear in the gradients of $\hat{\mathbf{n}}$, can emerge in the context of Hall currents and orbital magnetism in generic spin textures~\cite{Lux2020}. The corresponding chiral Hall effect (CHE) was shown to be directly sensitive to the sense of local spin chirality and, in contrast to the topological Hall effect, to fine details of spin distribution~\cite{Lux2020,Redies2020}, which potentially makes it a powerful tool for tracking texture dynamics with magnetotransport means. Recently, it was also demonstrated from model considerations and first-principles calculations that the CHE can be prominent in canted spin-orbit coupled ferromagnets and antiferromagnets~\cite{Kipp2021chiralComPhys}. For textures, a signal consistent with such chiral contributions has been reported experimentally for the anomalous Hall effect (AHE)~\cite{Monchesky2014MnSi}, the planar Hall effect (PHE)~\cite{Tokura2020PHE}, and magnetoconductivity (MC), although additional analysis is necessary to unambiguously pin down the exact microscopic origin of the observed signal.

In the context of chiral contributions to the magnetotransport of complex spin textures, understanding chiral transport properties of elemental one-dimensional spin-spiral states presents an important milestone and a pivotal starting point.
Spin-spirals ahve emerged as the ground state of various transition-metal compounds in bulk~\cite{Neubauer2009}, at surfaces~\cite{BodeNature2007}, and in one-dimensional systems~\cite{Menzel} as a result of complex exchange interactions.
Further, spin-spirals serve as natural building blocks for more complex spin structures such as multi-$q$ states~\cite{KubetzkaPRL2020}, domain walls~\cite{FreimuthSeemann2012} and skyrmions~\cite{Neubauer2009,HeinzeNature2011}, while their treatment often provides a key to modeling fluctuating magnets~\cite{Freimuth2020MagnonicSOT,Lichuan2020}.  Proper theoretical understanding of symmetry and microscopic mechanisms behind the transport properties of spin-spiral states thus provides a necessary foundation for building a coherent picture of transport phenomena in diverse spin textures. 
A distinct transport signal of spiral states has been observed in experiments lately~\cite{Monchesky2014MnSi}, and a recent theoretical work based on an effective Rashba model predicted the emergence of the chiral Hall effect for specific types of spin-spiral states~\cite{Lux2020SeibergWitten}. At this point, a comprehensive picture of magnetotransport of spin-spirals grounded in a realistic electronic structure on a specific lattice has emerged as a necessary next step.
Technically, computing transport properties of extended spin-spiral states is very challenging. The spin-spiral order in combination with spin-orbit interaction is inconsistent with the generalized Bloch theorem~\cite{Sandratskii1986,SandratskiiJOPCM1991}, which necessitates the use of very large unit cells containing thousands of atoms. The formidable computational effort explains the noticeable lack of expansive studies in this direction~(see e.g.~\cite{Kelly2012,FreimuthSeemann2012}).

In our work, we study in detail the magnetotransport properties of spin-spiral states in two-dimensional  magnets, and we make predictions concerning the impact of these properties on chiral transport fingerprints of large skyrmions.
We refer to an effective tight-binding model of electrons on a honeycomb lattice for explicit calculations of transport properties from the electronic structure of spin-spiral states in the Kubo linear-response formalism. Inspecting components of the conductivity tensor (anti-)symmetrized with respect to chirality, we uncover chiral contributions to anisotropic magnetoconductivity (denoted below as MC), the planar Hall effect, and the anomalous Hall effect. Specifically, we find perfect agreement between numerical results from the Kubo formalism and symmetry-based predictions from the gradient expansion for the existence of chiral contributions as a function of the spin-spiral propagation direction and the type of spiral.
We address intrinsic and extrinsic origins of the considered effects, uncovering a non-trivial competition of disorder and Berry phase effects on the chiral Hall effect exhibited by spin-spirals and skyrmions. 
The strong and extremely non-trivial response of the chiral transport to the pitch of the spirals found here suggests that chiral magnetotransport serves as a unique marker of the fine details of the spin distribution in spin textures. Finally, we show how by extracting effective parameters from explicit calculations of spin-spirals the chiral transport properties of large-scale skyrmions can be predicted.    

\section{Approach}
\subsection{Model}
We investigate the existence and properties of chiral magnetotransport effects on a bipartite honeycomb lattice of magnetic spins. To model the electronic structure, we employ  an  effective two-dimensional lattice  tight-binding Hamiltonian (in the $xy$-plane) which reads:
\begin{equation}
\begin{split}
H = -t \sum\limits_{\langle ij \rangle\alpha}  c_{i\alpha}^\dagger c_{j\alpha}^{\phantom{\dagger}} &+ i \alpha_{\rm R}\sum\limits_{\langle ij \rangle\alpha \beta}  \hat{\mathbf e}_z \cdot (\boldsymbol{\sigma} \times {\mathbf d}_{ij})_{\alpha\beta}\, c_{i\alpha}^\dagger c_{j\beta}^{\phantom{\dagger}}\\
&+ \lambda_{\rm ex} \sum_{i\alpha \beta} (\hat{\mathbf s}_i\cdot \boldsymbol{\sigma})_{\alpha\beta}\, c_{i\alpha}^\dagger c_{i\beta}^{\phantom{\dagger}},
\end{split}
\label{eq:model}
\end{equation}
where $c_{i\alpha}^\dagger$ ($c_{i\alpha}^{\vphantom{\dagger}}$) denotes the creation (annihilation) of an electron with spin $\alpha$ at site $i$, $\langle ...\rangle$ restricts the sums to nearest neighbors, the unit vector $\mathbf d_{ij}$ points from $j$ to $i$, and $\boldsymbol{\sigma}$ stands for the vector of Pauli matrices. Besides the hopping with amplitude $t$,  \cref{eq:model} contains the Rashba spin-orbit coupling of strength $\alpha_\text{R}$ originating for example in the surface potential gradient perpendicular to the plane (i.e.\,along $\nuv{z}$). The remaining term in \cref{eq:model} is the local exchange term with $\lambda_{\rm ex}$ characterizing the strength of exchange splitting and $\hat{\mathbf{s}}_i$ stands for the direction of spin on site $i$.  Here, we work with the following parameters of the model: $t=1.0$\,eV, $\alpha_{\rm R}=0.4$\,eV, and $\lambda_{\rm ex}=1.4$\,eV, which corresponds to the case of a strongly spin-orbit coupled magnet.

\subsection{Parametrization of spin spirals}\label{ParamSpiral}
In this work, we impose magnetic spiral distribution of spins $\hat{\mathbf{s}}_i$ in the Hamiltonian~(\ref{eq:model}), and we study the transport properties of such states. 
We proceed by first defining a continuous, normalized vector field $\hatn(\vec{x})$ which is discretized on the lattice as
\begin{align}
    \nuv{s}_i \equiv \hatn( \vec{R}_i - \vec{R}_0 ),
\end{align}
where $\vec{R}_i$ corresponds to the real-space position of lattice site $i$, and $\vec{R}_0$ represents a choice for the origin of the continuous coordinate system.
General spin-spirals are one-dimensional, periodic patterns of spins, whose modulation is characterized by a single phase factor $\Psi( \vec{x} ) = \vec{q}\cdot \vec{x}$.
Explicitly, we first construct such a spin-spiral in the $x$-direction followed by a rotation $R^{\vec{e}_z}_{\phi_q}$ around the $z$-axis into the actual direction of the $\vec{q}$-vector, as characterized by its polar coordinate $\phi_q$:
\begin{equation}
    \hatn = R^{\vec{e}_z}_{\phi_q} R^{\vec{e}_r}_{\Psi} \hatn_0 .
    \label{eq:n_spiral}
\end{equation}
Here, $\hatn_0$ defines the initial orientation of $\hatn$ for $\Psi = 0$.
Following the direction of $\vec{q}$ in real space, the pattern describes a rotational motion around the axis defined by $\vec{e}_r$.
With $\hatn_0 =  \vec{e}_z$ fixed, three different cases are considered here, see \cref{Fig1}: the Bloch spiral $\vec{e}_r = \vec{e}_x$, the N\'eel spiral $\vec{e}_r = \vec{e}_y$ and a tilted conical N\'eel phase $\vec{e}_r = \sin \alpha ~\vec{e}_y + \cos \alpha ~ \vec{e}_z$ with cone angle $0 \leq \alpha \leq \pi/2$, interpolating between the ferromagnetic phase for $\alpha = 0$ and the N\'eel spiral for $\alpha=\pi/2$. 
All different cases are contained in the parametrization:
\begin{equation}
\vec{e}_r = 
\sin \beta \sin \alpha ~\vec{e}_x  +
\cos \beta \sin \alpha ~\vec{e}_y + \cos \alpha ~ \vec{e}_z,
\label{eq:n_cone}
\end{equation}
where the angle $0 \leq \beta \leq \pi/2$ can rotate from the tilted N\'eel phase ($\beta=0$) to a tilted Bloch phase ($\beta=\pi/2$).

The wavevector pitch $q = \| \vec{q} \|$ is the main tuning knob to adjust the magnetization pattern in this work. The limiting cases are the ferromagnetic (FM) pattern for $q=0$ and the antiferromagnetic pattern for $q=\pi$. 
The $q=0$ limit corresponds to the regular out of plane (along the $z$-axis) FM case. 
However, depending on the orientation of the rotation axis $\vec{e}_r$ in real space, the two spins in the AFM pattern are not antiparallel, but enclose a polar angle of $2\cdot\alpha$, twice the cone angle that the rotation axis $\vec{e}_r$ encloses with the $z$-axis. This can be understood by realizing that a phase $\pi$ between spins on neighboring atomic sites corresponds to opposite positions on the cone around the rotation axis $\vec{e}_r$, which has an opening angle of $2\cdot\alpha$.

\subsection{Gradient expansion}

Within the linear response theory, the conductivity tensor $\sigma_{\alpha \beta}$ describes the electric current response in the system which is linear in electric field $\vec{E}$ as $j^\alpha = \sigma_{\alpha\beta} E^\beta$.
In order to arrive at a way to categorize  different physical effects that are encapsulated in the conductivity tensor, we start by writing an asymptotic expansion in the gradients $\partial_i n_j$ for a smooth magnetization texture, i.e.,
\begin{equation}
    \sigma_{\alpha \beta} [\hatn]
    \! =  \! \braket{ 
    \sigma_{\alpha \beta}^\mathrm{col}(\hatn) + 
    \sigma_{\alpha \beta \gamma \delta}^\chi(\hatn)
    \partial_\gamma n_\delta + \mathcal{O}(\partial^2) }.
    \label{eq:gradient_expansion}
\end{equation}
Here, the bracket $\braket{\bullet}$ indicates the real-space integral $\int \dd\vec{x} / \mathcal{V}$, where $\mathcal{V}$ is the real-space volume of the system, $\vec{x}=(x,y)$ is the position vector, and notation $[\hatn]$ implies the functional dependence on the overall texture as given by the distribution of $\hatn$ in real space.
The conductivity tensor itself can be decomposed into symmetric and antisymmetric components with respect to the interchange of indices $\alpha$ and $\beta$:
$
      \sigma_{\alpha \beta} [\hatn] = 
        \sigma_{[\alpha \beta]} [\hatn] +   \sigma_{(\alpha \beta)} [\hatn]
$. 
Here $[\alpha\beta]$ indicates the antisymmetrization, whereas $(\alpha\beta)$ represents a symmetrization of indices.
Via the Onsager reciprocity relations, $ \sigma_{[\alpha \beta]} [\hatn] = -  \sigma_{[\alpha \beta]} [-\hatn]$ and $\sigma_{(\alpha \beta)} [\hatn] = \sigma_{(\alpha \beta)} [-\hatn]$.
The symmetric tensor then describes magnetoconductivity, anisotropic magnetoconductivity and planar Hall effects, while its antisymmetric counterpart captures the anomalous Hall effect. 
The superscript "col" indicates that $\sigma_{\alpha \beta}^\mathrm{col}(\hatn(\vec{x}))$ fully describes the linear response of a collinear magnetized state. 
For slowly varying textures, the next-to-leading-order term $\sigma_{\alpha \beta \gamma \delta}^\chi(\hatn(\vec{x}))$ couples to the first-order gradients of the magnetization texture and is therefore sensitive to the chirality of $\hatn$, thereby providing the {\it chiral} part of the overall conductivity tensor.  When the texture constitutes a spin-spiral with a wave-vector $\mathbf{q}$, as introduced above, the corresponding chiral corrections to the conductivity are fully antisymmetric in $\mathbf{q}$.

\begin{table}[t!]
	\centering
	\caption{{\bf Character table of $C_{6v}$.} 
	Shown are the characters of each irreducible representation for each conjugacy class, alongside linear and quadratic basis functions which generate the respective representations.}
	\label{tab:character_table}\vspace{0.2cm}
	\begin{tabular}{c|cccccc|c|c}
	  $C_{6v}$ & $E$ & 2$C_6\text{(z)}$ & $2C_3\text{(z)}$ & $C_2\text{(z)}$ & 3$\sigma _v$ & 3$\sigma _d$ & linear & quadratic \\ \toprule
 $A_1$ & 1 & 1 & 1 & 1 & 1 & 1   &  $z$ & $x^2+y^2, z^2$ \\
 $A_2$ & 1 & 1 & 1 & 1 & -1 & -1& $n_z$ & \\
 $B_1$ & 1 & -1 & 1 & -1 & 1 & -1 & &\\
 $B_2$ & 1 & -1 & 1 & -1 & -1 & 1& & \\
 $E_1$ & 2 & 1 & -1 & -2 & 0 & 0 &$(x,y)$ & $(xz, yz)$\\
 $E_2$ & 2 & -1 & -1 & 2 & 0 & 0&  & $(x^2 -y^2, xy)$\\ \bottomrule
	\end{tabular}
\end{table}
\begin{table}[t!]
	\centering
	\caption{{\bf Kronecker products in $C_{6v}$.} Shown is the reduction of all possible combinations of Kronecker products among the irreducible representations of $C_{6v}$. }
	\label{tab:tensor_producs}\vspace{0.2cm}
	\begin{tabular}{c|cccccc}
	 $\otimes$  & $A_1$ & $A_2$ & $B_1$ & $B_2$ & $E_1$ & $E_2$ \\ \toprule
 $A_1$ & $A_1$ & $A_2$ & $B_1$ & $B_2$ & $E_1$ & $E_2$ \\
 $A_2$ & $A_2$ & $A_1$ & $B_2$ & $B_1$ & $E_1$ & $E_2$ \\
 $B_1$ & $B_1$ & $B_2$ & $A_1$ & $A_2$ & $E_2$ & $E_1$ \\
 $B_2$ & $B_2$ & $B_1$ & $A_2$ & $A_1$ & $E_2$ & $E_1$ \\
 $E_1$ & $E_1$ & $E_1$ & $E_2$ & $E_2$ & $A_1+A_2+E_2$ & $B_1+B_2+E_1$ \\
 $E_2$ & $E_2$ & $E_2$ & $E_1$ & $E_1$ & $B_1+B_2+E_1$ & $A_1+A_2+E_2$ \\ \bottomrule
	\end{tabular}\\
\end{table}

\begin{table*}[t!]
	\centering
	\caption{{\bf Classification of electric transport effects in a noncollinear magnet.} 
	For the symmetry group $C_{6v}$, the conductivity tensor in the $x$-$y$--plane factors into the irreducible representations $\sigma = E_1 \otimes E_1 =  A_1 + A_2 + E_2$.
	Each representation is generated from specific combinations of the components $\sigma_{\alpha\beta}$ which we refer to as channels.
	Within each channel, different kinds of physical effects can appear which are sensitive to different aspects of the underlying magnetization texture and which we roughly divide into those effects which are already present in a collinear ferromagnet and those which require a finite, first-order gradient $\partial_i n_j$.
	}
	\label{tab:effects}\vspace{0.2cm}
	\begin{tabular}{c|cp{2cm}p{6cm}p{6cm}}
	 Irrep  & Channel & Name & Collinear effects $ \mathcal{O}(\partial^0) $  & Chiral effects $ \mathcal{O}(\partial^1) $ \\ \toprule
	 $A_1$  & $(\sigma_{(xx)} + \sigma_{(yy)} ) / 2$ & isotropic longitudinal & longitudinal conductivity (LC), magnetoconductivity (MC), anisotropic magnetoconductivity (AMC) &  chiral  magnetoconductivity (CMC)\\
	 $A_2$  & $\sigma_{[xy]}$ & antisymmetric transverse & anomalous Hall effect (AHE) & chiral Hall effect (CHE) \\
	 $(E_2)_1 $  & $(\sigma_{(xx)} - \sigma_{(yy)} ) / 2$ & anisotropic longitudinal & longitudinal planar Hall effect (LPHE) & chiral longitudinal planar Hall effect (CLPHE)  \\
	 $(E_2)_2 $  & $\sigma_{(xy)}$ & symmetric transverse & planar Hall effect (PHE) & chiral planar Hall effect (CPHE) 
\\ \bottomrule
	\end{tabular}\\
\end{table*}

Following \cite{Birss1964}, one can expand $\sigma^\mathrm{col}$ and $\sigma^\chi$ into powers of the magnetization vector $\hatn$ and reduce the number of possible coupling terms using the restrictions imposed by the crystallographic point group of the nonmagnetic lattice.  
For example, this method has been successfully applied recently to study the chiral corrections to the spin-Hall magnetoresistance in the noncollinear magnet \ce{Cu2OSeO3} \cite{MostovoyPRB2021}.
Below, we demonstrate the way that it can be done for the partial case of $C_{6v}$ symmetry of the Hamiltonian as given by Eq.~\eqref{eq:model}.
A systematic way to perform this reduction is guided by the representation theory of $C_{6v}$ as summarized by the character table in Tab.~\ref{tab:character_table}.
Accordingly, the gradient operator $\nabla = ( \partial_x, \partial_y)$ generates the representations $\Gamma_{\vec{x}} = \Gamma_{\nabla}=E_1$.
The magnetization $\hatn$ -- as an axial vector -- decomposes into $\hatn_\parallel = (n_x, n_y,0)$ belonging to $E_1$ and $\hatn_\perp = (0,0,n_z) $, belonging to $A_2$, i.e., $\Gamma_{\hatn} = A_2 + E_1$.
Using the Schur orthogonality relations~\cite{dresselhaus2007group,hamermesh2012group}, one can derive the decomposition of the Kronecker products $\Gamma_i \otimes \Gamma_j = \bigoplus_{k} \lambda_k \Gamma_k$ among any of the irreducible representations (irreps) $\Gamma_i$, where $\lambda_k$ are positive integers. 
Summarized in Tab.~\ref{tab:tensor_producs}, the result of this procedure can be used to study the symmetry-allowed couplings in the expansion of $\sigma_{\alpha\beta}$: since it couples two polar vectors in the $x$-$y$--plane, it carries the representation $\sigma = E_1 \otimes E_1 =  A_1 + A_2 + E_2$.
Different irreps therefore distinguish different categories of physical effects which are summarized in Table \ref{tab:effects} and which will be introduced in the following.
Due to the Onsager relations, the $A_1$ and $E_2$ contributions to $\sigma_{\alpha\beta}$ have to be even in $n_i$ while the $A_2$ term is odd.
In the absence of magnetism or possible external fields, all components that do not belong to the totally symmetric representation $A_1$ have to vanish by Neumann's symmetry principle~\cite{neumann1885vorlesungen, nowick2005crystal}.
The $A_1$ term $ (\sigma_{(xx)} + \sigma_{(yy)} ) /2$ then captures the isotropic contributions to the longitudinal conductivity.
In the presence of a finite magnetization, the components of $\hatn$ and its gradients can form a basis for the irreducible representations $A_2$ and $E_2$, thereby leading to a finite anomalous Hall effect in $ \sigma_{[xy]} $ and a finite planar Hall effect in $\sigma_{(xy)}$ for example.
Using representation theory, one can study what different terms appear in the gradient expansion of Eq.~(\ref{eq:gradient_expansion}) if each of the tensors is expanded as a power series in $n_i$.
For the collinear case, one finds
that the magnetization carries the following irreducible representations
\begin{align}
    A_1 \colon &  n_\parallel^2, n_\perp^2
    \\
    A_2 \colon &  n_z
    \\
    E_1 \colon &  (n_x n_z, n_y n_z )
    \\
    E_2 \colon &  \left ( ( n_x^2 - n_y^2 )/2 , n_x n_y  \right) ,
\end{align}
up to second order in $n_i$, where $n_\perp^2 = n_z^2$ and $n_\parallel^2 = n_x^2 + n_y^2$.
Here, one has to take into account that $\hatn$ transforms as a pseudovector.
At the given order in $n_i$, there are thus $5$ couplings to the collinear magnetic state:
\begin{align}
    \sigma_{A_1}^\mathrm{col}(\hatn)
    &= \gLR + \gMR \| \hatn \|^2 + \gAMR (n_\perp^2- n_\parallel^2) + \mathcal{O}(n_i^4)
    \\
    \sigma_{A_2}^\mathrm{col}(\hatn)
    &= \gAHE n_z + \mathcal{O}(n_i^3)
    \\
    \sigma_{E_2}^\mathrm{col}(\hatn)
    &= \gPHE ( (n_x^2-n_y^2)/2, n_x n_y ) + \mathcal{O}(n_i^4),
\end{align}
which we refer to as the longitudinal conductivity (LC), the magnetoconductivity (MC), the anisotropic magnetoconductivity (AMC), the anomalous Hall effect (AHE) and the planar Hall effect (PHE).
The PHE is commonly understood as the off-diagonal components $\sigma_{(xy)}$ of the symmetrized conductivity tensor. 
Due to the peculiarity of the $C_{6v}$ - symmetry, the same coefficient controls anisotropic contributions to the longitudinal conductivity $ (\sigma_{(xx)} - \sigma_{(yy)} ) /2$. 
In Table \ref{tab:effects}, this latter term is therefore listed as the longitudinal planar Hall effect (LPHE).

Moving on to the gradient-induced effects in two dimensions, one can deduce from Tab.~\ref{tab:character_table} and Tab.~\ref{tab:tensor_producs} that $\Gamma_{\nabla \otimes \hatn} \equiv E_1 \otimes (A_2 + E_2) =A_1 + A_2 + E_1 + E_2$. 
The relevant irrep is $A_2$ (due to the Onsager relations), and it occurs only once:
\begin{align}
A_1 \colon & (\nabla \times \hatn)_z
\\
A_2 \colon & \nabla \cdot \hatn.
\\
E_1 \colon & ( \partial_x n_z ,  \partial_y n_z )
\\
E_2 \colon &  ( \partial_x n_x - \partial_y n_y, \partial_x  n_y + \partial_y n_x ).
\end{align}
We assume that partial integration can be performed under the real-space integral in Eq.~(\ref{eq:gradient_expansion}) which renders the $A_2$ contribution  integrate to zero.
The next order in $n_i$ can be obtained by reducing the tensor product 
$\Gamma_{\nabla \otimes \hatn}\otimes \Gamma_{\hatn} = 2 A_1+2 A_2+B_1+B_2+4 E_1+2 E_2$:
\begin{align}
    A_2 \otimes A_2 \to A_1\colon & 
    (\nabla \cdot \hatn) n_z
    \\
    E_1\otimes A_2 \to A_1\colon & 
   (\hatn \cdot \nabla) n_z
    \\
   E_2\otimes A_2 \to E_2 \colon &
   n_z ( \partial_x n_x - \partial_y n_y, \partial_x  n_y + \partial_y n_x ), 
    \\
   E_1\otimes E_1 \to E_2 \colon &  ( n_x \partial_x n_z - n_y \partial_y n_z ,  n_x \partial_y n_z  + n_y \partial_x n_z ).
\end{align}
Under partial integration, one recognizes that the $A_1$ contributions combine into the Lifshitz invariant $(\nabla \cdot \hatn)  n_z - (\hatn \cdot \nabla) n_z$ which is also responsible for the DMI interaction under $C_{6 v}$ symmetry~\cite{dzyaloshinskii1964theory,bogdanov1989thermodynamically}.
The two $E_2$ contributions are the same under partial integration as well.
Combined, this leads to
\begin{align}
    \sigma_{A_1, \alpha \beta}^\chi(\hatn) \partial_\alpha n_\beta 
    =& \gCMC  (\nabla \cdot \hatn- \hatn \cdot \nabla)  n_z  + \mathcal{O}(n_i^4),
    \\ 
    \sigma_{E_2, \alpha \beta}^\chi(\hatn) \partial_\alpha n_\beta 
    = &\gCPHE~n_z ( \partial_x n_x - \partial_y n_y, \partial_x  n_y + \partial_y n_x )
    \notag \\ & + \mathcal{O}(n_i^4),
\end{align}
describing the chiral magnetoconductivity (CMC) and the chiral planar Hall effect (CPHE).

The symmetrized tensor product $\mathrm{Sym}( \hatn \otimes \hatn)$ carries the representation $\Gamma_{\mathrm{Sym}( \hatn \otimes \hatn)}= 2 A_1 + E_1 +E_2 $. 
For the next order in $n_i$, one therefore finds
$\Gamma_{\nabla \otimes \hatn}\otimes \Gamma_{\mathrm{Sym}( \hatn \otimes \hatn)} = 4 A_1+4 A_2+2 B_1+2 B_2+6 E_1+6 E_2$. 
Since this order is odd under time reversal, we are only interested in the four $A_2$ representations given by
\begin{align}
    A_2 \otimes A_1 \to 2A_2\colon & 
    (\nabla \cdot \hatn ) n_\perp^2, (\nabla \cdot \hatn ) n_\parallel^2, \\
     E_1 \otimes E_1 \to A_2\colon &  ( \hatn \cdot \nabla ) n_\perp^2 /2
     \\
     E_2 \otimes E_2 \to A_2\colon & n_x n_y (\partial_x n_y + \partial_y n_x) \\ & + (n_x^2-n_y^2) ( \partial_x n_x - \partial_y n_y )/2 .
\end{align}
The last term may be recombined with other $A_2$ representations of this order such that it can be written as $( \hatn \cdot \nabla ) n_\parallel^2 /2$.
And therefore
\begin{align}
    \sigma_{A_2, \alpha \beta}^\chi(\hatn) \partial_\alpha n_\beta 
    =& \gCHE ~(\nabla \cdot \hatn ) ( n_\perp^2 - n_\parallel^2 )  + \mathcal{O}(n_i^5) ,
\end{align}
is the only coupling that describes the chiral Hall effect (CHE) in $C_{6v}$ symmetric systems.
In summary, we find that
\begin{align}
\sigma_{A_1}[\hatn] =& (\gLR + \gMR) + \gAMR \braket{n_\perp^2 - n_\parallel^2}
\notag \\ &
     + \gCMC\braket{ (\nabla \cdot \hatn- \hatn \cdot \nabla)  n_z }
     + \mathcal{O}(n_i^4)
\label{eq:local_results_1}
\\
\sigma_{A_2}[\hatn] =& \gAHE \braket{ n_z } + \gCHE  \braket{(\nabla \cdot \hatn ) ( n_\perp^2 - n_\parallel^2 )} + \mathcal{O}(n_i^5) 
\label{eq:local_results_2}
\\
\sigma_{E_2}[\hatn] =&\gPHE \braket{  ( (n_x^2-n_y^2)/2, n_x n_y ) }
\notag \\ & + \gCPHE  \braket{ n_z ( \partial_x n_x - \partial_y n_y, \partial_x  n_y + \partial_y n_x ) }
\notag \\ &+ \mathcal{O}(n_i^4), 
\label{eq:local_results_3}
\end{align}
where the factors $\gamma_i$ are temperature-dependent material constants that can be extracted from an underlying microscopic model.
Inserting the spin-spiral defined by the combination of Eq.~(\ref{eq:n_spiral}) and Eq.~(\ref{eq:n_cone}), one arrives at
\begin{align}
\sigma_{A_1}[\hatn] =& (\gLR + \gMR) + \frac{1}{2} \gAMR \cos ^2(\alpha ) (3 \cos (2 \alpha )-1)
\notag \\ &
+ \gCMC q  \cos \beta \sin^3 \alpha
     + \mathcal{O}(n_i^4)
\\
\sigma_{A_2}[\hatn] =& \gAHE \cos^2 \alpha +  \frac{1}{2} \gCHE  q \cos \beta  \sin (\alpha ) \sin ^2(2 \alpha ) 
\notag \\ &
+ \mathcal{O}(n_i^5) 
\\
\sigma_{E_2}[\hatn] =& -\frac{1}{8}\gPHE \sin ^2(\alpha ) (3 \cos (2 \alpha )+1)  \boldsymbol{\vartheta}_2
\notag \\ & + 
\frac{1}{2} \gCPHE q \sin ^3(\alpha ) \boldsymbol{\vartheta}_1 + \mathcal{O}(n_i^4) ,
\end{align}
where  $\boldsymbol{\vartheta}_n   = ( \cos(n\beta-2\phi_q), -\sin(n \beta - 2 \phi_q))$.

\begin{table}[t!]
	\centering
	\caption{{\bf The existence of chiral magnetotransport of spin-spiral states.} For each type of the spiral state (N\'eel, Bloch, cone) the emergence of the corresponding effect [non-chiral (AHE) and chiral (CHE) anomalous Hall effect, non-chiral (MC) and chiral (CMC) magnetoconductivity, non-chiral (PHE) and chiral (CPHE) chiral planar Hall effect] is marked with the sign ``$\checkmark$" when it is allowed by symmetry arguments of the gradient expansion and confirmed by explicit Kubo transport calculations, while its absence is marked with an empty entry. Crystallographic directions $[100]$, $[0\overline{1}1]$ and $[110]$ mark the direction of the $\mathbf{q}$-vector, which correspond to the value of $\phi_q$ of 0, $-\pi/2$ and $\pi/3$, respectively.}
	\label{tab:TypesHE}\vspace{0.2cm}
	\begin{tabular}{|c|l|c|c||c|c||c|c|}
		\hline
		\multicolumn{2}{|c|}{Type}&AHE&CHE&MC&CMC&PHE&CPHE\\ 
		\hline\hline
		\multirow{3}*{N\'eel}&[100]  &&&\checkmark&\checkmark&&\\\cline{2-8}
		&[0$\overline{1}$1] &&&\checkmark&\checkmark&&\\\cline{2-8}
		&[110] &&&\checkmark&\checkmark&\checkmark&\checkmark\\
		\hline\hline
		\multirow{3}*{Bloch}&[100] &&&\checkmark&&&\checkmark\\\cline{2-8}
		&[0$\overline{1}$1] &&&\checkmark&&&\checkmark\\\cline{2-8}
		&[110] &&&\checkmark&\checkmark&\checkmark&\checkmark\\
		\hline\hline
		\multirow{3}*{Cone}&[100] &\checkmark&\checkmark&\checkmark&\checkmark&&\\\cline{2-8}
		&[0$\overline{1}$1] &\checkmark&\checkmark&\checkmark&\checkmark&&\\\cline{2-8}
		&[110] &\checkmark&\checkmark&\checkmark&\checkmark&\checkmark&\checkmark\\
		\hline
	\end{tabular}\\
\end{table}

\begin{figure*}[ht!]
	    \includegraphics[width=1.0\hsize]{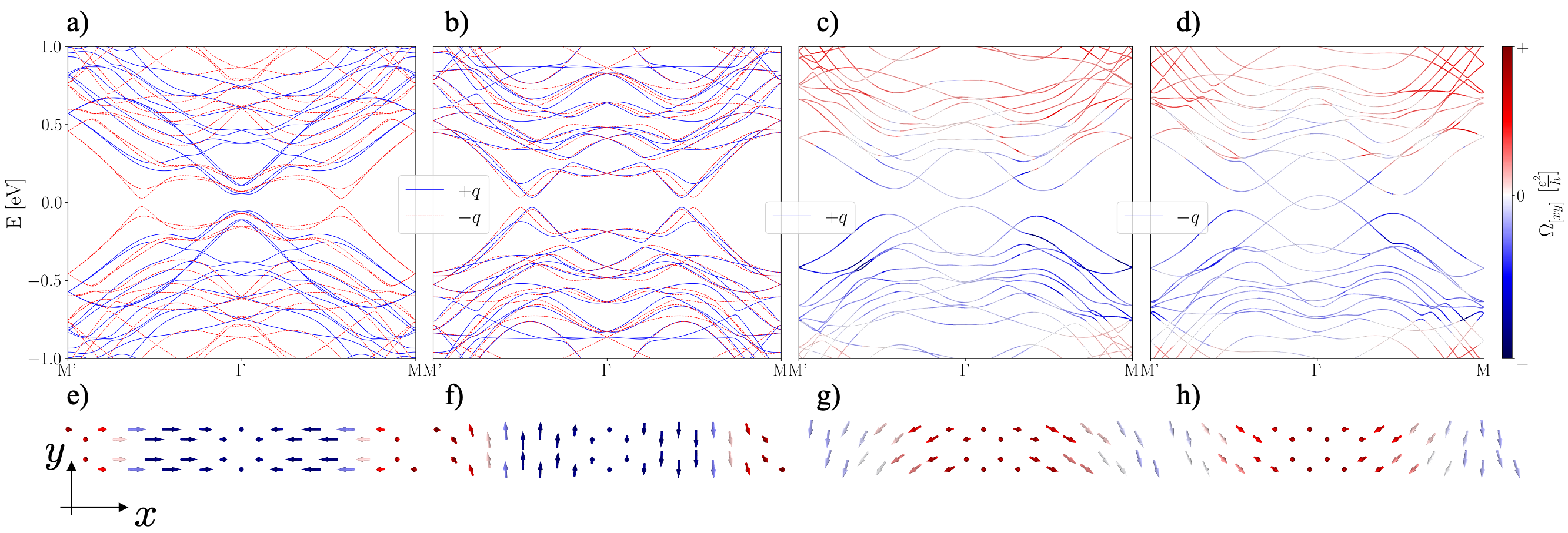}
		\caption{{\bf Electronic structure of N\'eel, Bloch and cone-type spirals.} 
		 Bandstructures of N\'eel (a) and Bloch (b) spin-spirals in the energy interval $[-1,1]$ eV for $\vec{q}=\pm0.31a_0^{-1}\cdot\nuv{x}$ (blue and red).
		 (c-d) Bandstructures of a cone-type spiral with the cone angle $\alpha = \SI{0.6}{rad}\approx\ang{34.38}$ in the energy range of $[-1,1]$\,eV for opposite orientations of the wavevector $\vec{q}=\pm 0.31a_0^{-1}\cdot\nuv{x}$,  reflecting  the changes in the electronic structure due to opposite sense of chirality. The color of the bands in (c-d) indicates the value of the band-resolved Berry curvature $\Omega_{xy}(\vec{k})$. (e-h) The real-space magnetic texture corresponding to  band structures in the respective columns. For more details see main text.
		}
		\label{Fig2Z}
\end{figure*}

\subsection{Kubo Formalism}
Given specific electronic structure, we calculate the transverse and diagonal conductivity at zero temperature using the Kubo formalism, which allows us to take into account the effect of disorder in the system on the conductivity tensor. In order to do so,  we replace the retarded and advanced Green functions $G_0$ of the perfect crystal by the full Green function $G = \frac{1}{G_0^{-1}-\Sigma}$, where $\Sigma(E,\vec{k})$ is the self-energy representing the effect of disorder. Within this work, we are using a constant broadening model such that $\Sigma(E,\vec{k})=- i\Gamma\cdot \myMat{I}$.
With the constant broadening $\Gamma$ we obtain a Green function diagonal in the eigenspace of the Hamiltonian:
\begin{align}
    G^{R/A}(E,\vec{k})_{mn}=\frac{\delta_{mn}}{E-\epsilon_{n\vec{k}}\pm i\Gamma},
\end{align}
where $\epsilon_{n\vec{k}}$ are the single-electron eigenenergies. The antisymmetric part of the conductivity tensor, which can be expressed in terms of $G^{R/A}$~\cite{CzajaAHE}, splits into two contributions:
\begin{align}
    \sigma_{[\alpha\beta]}^{I}=&-\frac{1}{2\pi}\int\frac{d^3k}{(2\pi)^3}\sum_{\substack{mn\\m\neq n}}\Im\lbrace v_{mn}^{\alpha}(\vec{k})v_{nm}^{\beta}(\vec{k})\rbrace\\
\nonumber    &\times \frac{(\epsilon_{m\vec{k}}-\epsilon_{n\vec{k}})\Gamma}{((E_F -\epsilon_{m\vec{k}})^2+\Gamma^2)((E_F -\epsilon_{n\vec{k}})^2+\Gamma^2)}
\end{align}
and
\begin{align}
    \sigma_{[\alpha\beta]}^{II}=&-\frac{1}{\pi}\int\frac{d^3k}{(2\pi)^3}\sum_{\substack{mn\\m\neq n}}\Im\lbrace v_{mn}^{\alpha}(\vec{k})v_{nm}^{\beta}(\vec{k})\rbrace\\
\nonumber        &\times         \frac{\Gamma}{(\epsilon_{m\vec{k}}-\epsilon_{n\vec{k}})((E_F -\epsilon_{m\vec{k}})^2+\Gamma^2)}\\
\nonumber        &-\frac{1}{(\epsilon_{m\vec{k}}-\epsilon_{n\vec{k}})^2}
        \Im\left\lbrace\ln{\left(\frac{E_F-\epsilon_{m\vec{k}}+i\Gamma}{E_F-\epsilon_{m\vec{k}}+i\Gamma}\right)}\right\rbrace,
\end{align}
where $\alpha$ and $\beta$ are the Cartesian indices and $E_F$ is the Fermi energy.
We refer to $\sigma_{\alpha\beta}^{I}$ as the Fermi-surface term, since it only picks up contributions from the Fermi surface. The term $\sigma_{\alpha\beta}^{II}$ collects terms from all occupied states up to the Fermi level and is therefore referred to as the Fermi-sea term.
The symmetric part of the conductivity tensor is given by~\cite{CzajaThesis}
\begin{align}
           \sigma_{(\alpha\beta)}=&\frac{1}{\pi}\int\frac{d^3k}{(2\pi)^3}
           \sum_{mn}\Re\lbrace v_{mn}^{\alpha}(\vec{k})v_{nm}^{\beta}(\vec{k})\rbrace\\
\nonumber   &\times \frac{\Gamma^2}{((E_F - \epsilon_{m\vec{k}})^2+\Gamma^2)((E_F - \epsilon_{n\vec{k}})^2+\Gamma^2)}.
\end{align}
In evaluating the Kubo expressions for the conductivity of spin-spiral states we have considered systems treated in a super-cell with up to 1600 atoms in the unit cell, using up to $2\cdot 10^5$ $k$-points for performing Brillouin zone integrations.

\section{Chiral transport properties of spin-spirals}\label{Results}
Below we compare the results of the gradient expansion performed up to linear order for the considered model to the explicit tight-binding calculations of the conductivity tensor by using the Kubo formalism.
From explicit calculations of the conductivity for the system in a spin-spiral state $\hatn_\mathrm{cone}$  (defined in Eq.~(\ref{eq:n_spiral}) and Eq.~(\ref{eq:n_cone})), we extract the contributions to the conductivity tensor which are even (non-chiral, $\sigma^{nc}$) and odd (chiral, $\sigma^{c}$) in spiral wave-vector $\vec{q}$ by performing the corresponding decomposition: 
\begin{align}
    \sigma_{\alpha\beta}^{c(nc)} = \frac{\sigma_{\alpha\beta}(\vec{q})\mp\sigma_{\alpha\beta}(-\vec{q})}{2}.
\end{align}
Using this definition, we can make a connection to the gradient expansion in the long-wavelength limit.
Here, one finds the asymptotic relationships
\begin{align}
    \sigma_{\alpha\beta}^{nc} 
    & \sim \braket{\sigma^\mathrm{col}_{\alpha\beta} (\hatn) },   ~ \text{for} ~ q \to 0,
    \\ 
    \sigma_{\alpha\beta}^{c}  &\sim 
    \braket{\sigma^\chi_{\alpha\beta \gamma \delta} (\hatn) \partial_\gamma n_{\delta} }
   , ~ \text{for} ~ q \to 0.
\end{align}
We thus scrutinize the existence of chiral contributions to the magnetoconductivity, $\sigma^{c}_{(\alpha\alpha)}$ ($\vec{q}$-chiral part of the diagonal components of the conductivity tensor $\sigma_{\alpha\alpha}$), chiral planar Hall effect, $\sigma^{c}_{(\alpha\beta)}$  ($\vec{q}$-chiral part of the $\alpha\leftrightarrow\beta$ symmetric off-diagonal components of the conductivity tensor $\sigma_{\alpha\beta}$) and chiral Hall effect, $\sigma^{c}_{[\alpha\beta]}$  ($\vec{q}$-chiral part of the $\alpha\leftrightarrow\beta$ antisymmetric off-diagonal components of the conductivity tensor $\sigma_{\alpha\beta}$).
The predictions of the explicit calculations concerning the existence of MC and CMC, PHE and CPHE, AHE and CHE, are presented in~\cref{tab:TypesHE}. They are entirely consistent with the symmetry analysis of the gradient expansion also beyond the long-wavelength limit: whenever an empty instance is met in the table, the gradient expansion predicts a vanishing contribution for a given direction of the spin-spiral and its type, while explicit calculations provide negligible values of the conductivity. Below, we discuss in detail the emergence of chiral contributions to the MC, PHE and AHE.

\begin{figure*}[t!]
	    \includegraphics[width=\hsize]{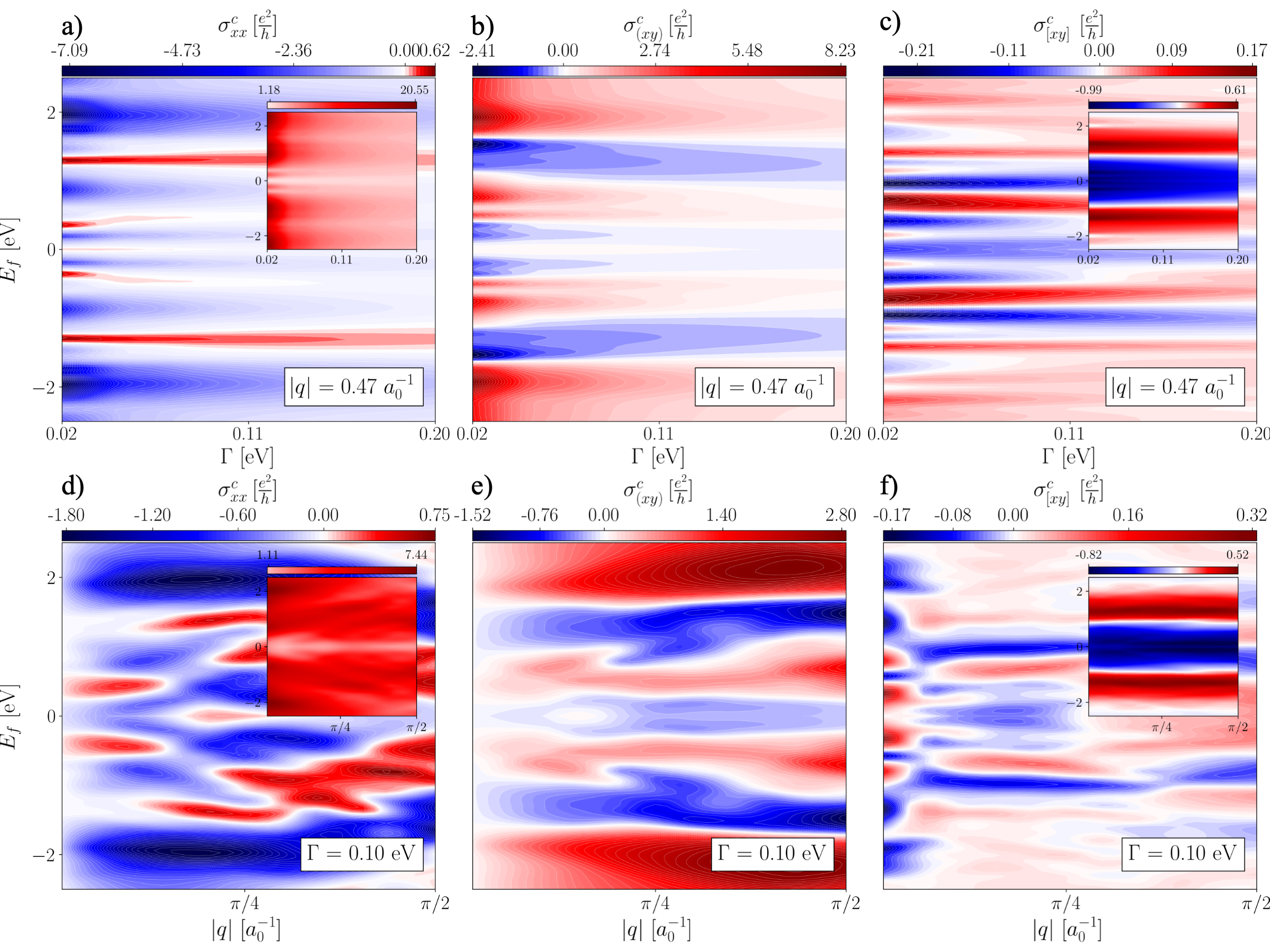}
		\caption{{\bf Chiral magnetoconductivity (CMC), chiral planar Hall effect (CPHE) and chiral Hall effect (CHE).} (a,d):  chiral magnetoconductivity (CMC), $\sigma^c_{xx}$, calculated for a N\'eel-type spiral with $\vec{q}=\pm 0.47a_0^{-1}\cdot\nuv{x}$ as a function of Fermi energy $E_f$ and broadening $\Gamma$, (a), and as a function of $E_f$ and $q$ (along $\nuv{x}$) at $\Gamma$ of 100\,meV, (d). The insets depict the corresponding behavior of the non-chiral part of the conductivity tensor element, $\sigma^{nc}_{xx}$. (b,e): 
		chiral planar Hall effect (CPHE), $\sigma^c_{(xy)}$, calculated for a Bloch spiral with $\vec{q}=\pm 0.47a_0^{-1}\cdot\nuv{x}$ as a function of $E_f$ and $\Gamma$, (b), and as a function of $E_f$ and $q$ (along $\nuv{x}$) at $\Gamma$ of 100\,meV, (e). The corresponding non-chiral part of the conductivity tensor element, $\sigma^{nc}_{(xx)}$, vanishes in this case, see \cref{tab:TypesHE}.
		(c,f): chiral Hall effect (CHE), $\sigma^c_{[xy]}$, calculated for a cone-type spiral with $\vec{q}=\pm 0.47a_0^{-1}\cdot\nuv{x}$ and a cone angle of $\alpha = \SI{0.6}{rad}\approx\ang{34.38}$ as a function of $E_f$ and $\Gamma$, (c), and as a function of $E_f$ and $q$ (along $\nuv{x}$) at $\Gamma$ of 100\,meV, (f). The insets depict the corresponding behavior of the non-chiral part of the conductivity tensor element, $\sigma^{nc}_{[xy]}$.
		}
		\label{Fig3}
\end{figure*}

\subsection{Longitudinal chiral conductivity}
The discussion of chiral effects in the longitudinal conductivity falls into two categories: the isotropic contributions $ (\sigma_{(xx)} + \sigma_{(yy)} ) / 2$ from the totally symmetric irrep $A_1$, which we defined as CMC, and anisotropic contributions $ (\sigma_{(xx)} - \sigma_{(yy)} ) / 2$ from irrep $E_2$, which we referred to as CLPHE.
For the spiral, the two effects evaluate to 
\begin{align}\label{LongRes}
    \gCMC\braket{ (\nabla \cdot \hatn- \hatn \cdot \nabla)  n_z } =& \gCMC q  \cos \beta \sin^3 \alpha,
    \\
    \gCPHE  \braket{ n_z ( \partial_x n_x - \partial_y n_y ) }
    =&\frac{1}{2} \gCPHE q \sin ^3(\alpha ) \cos(\beta- 2 \phi_q ) .
\end{align}
For a Bloch-type spiral, $\beta=\pi/2$, the CMC is zero and the CLPHE terms only exist along specific directions: $\cos(\pi/2 - 2 \phi_q ) = \sin(2 \phi_q)$. From this argument it is clear that a Bloch spiral along the $x$-direction will not display any chiral longitudinal conductivity.
Consequently, we take these longitudinal effects under closer inspection for a N\'eel-type spiral, $\beta=0$. Inspecting \cref{LongRes} we realize that
\begin{equation}
  \sigma_{(xx)}^c= \sigma_{A_1} + \sigma_{E_2}^1
\end{equation}
contains both isotropic and anisotropic components. However, both isotropic and anisotropic components follow the angular dependence
dictated by symmetry
\begin{equation}
  \sigma_{(xx)}^c(\alpha) = \sin^3 \alpha \left. \sigma_{(xx)}^c \right|_{\alpha = \pi/2} 
  \label{eq:cos3}
\end{equation}
in the long-wavelength limit, independent of the type of spiral.

The bandstructure of the model for the N\'eel spiral  with $\pm\mathbf{q}$ along $x$ and the magnitude of $|\vec{q}|=0.31\,a_0^{-1}$, presented in \cref{Fig2Z}~a), shows a strong chiral asymmetry. Drastic modifications brought to the energetic positions of the states as well as their relative splittings give rise to a pronounced chiral contribution to the longitudinal conductivity.
In \cref{Fig3}~(a) we show the CMC-signal given by $\sigma_{xx}^{c}$ as a function of Fermi energy $E_f$ and broadening $\Gamma$, taking a specific value of $|\vec{q}| = 0.47\,a_0^{-1}$ for a N\'eel spiral propagating along the $x$-axis. For comparison, we show the non-chiral MC in an inset.
We observe the particle-hole symmetry of the model which is evident from the symmetry of $\sigma_{xx}^{c}$ with respect to Fermi energy $E_f$. This symmetry is fulfilled by all quantities displayed in \cref{Fig3}~(a-f).   
Strong contributions to CMC, shown in  \cref{Fig3}~(a), arise from certain specific energies (e.g.~$\pm\SI{1.3}{eV}$ and broader humps around $\pm\SI{2.0}{eV}$), corresponding to the features in the electronic structure which are most affected by chirality. In contrast, non-chiral MC given by $\sigma_{xx}^{nc}$, is less sensitive to the Fermi energy. The magnitude of the computed CMC signal reaches as much as 30\% of the non-chiral part of the conductivity. Clearly, the largest values of CMC are reached for the smallest values of $\Gamma$, and the rapid decay of $\sigma_{xx}^{c}$ with $\Gamma$ is evident. A careful analysis of the CMC scaling with $\Gamma$ for very small values of $\Gamma$ reveals an expected $1/\Gamma$ behavior. 

In \Cref{eq:cos3} we uncovered a direct connection between the CMC for a N\'eel spiral and that of the cone-type spiral, valid in the limit of small spin orbit coupling (SOC). This connection is intuitive in the sense that the N\'eel spiral is the limiting case of $\beta=0$ for the cone-type spiral, where the cone angle approaches $\alpha=\pi/2$. We examine the prediction of \Cref{eq:cos3} made from the relation between the N\'eel and the cone-type spiral in \Cref{Fig4}~a),
comparing the scaling of the CMC-signal for the cone-type spiral, $\sigma_{xx,\text{Cone}}^{c}$ (blue), to the function $\sin^3(\alpha)\cdot\sigma_{xx,\text{Neel}}^{c}$ (red). For SOC strength of up to $\alpha_R=\SI{0.04}{eV}$ we observe a perfect match where the cone-type spiral is equivalent to a N\'eel spiral (left side, $\alpha=0$) up to cone angles as large as $\pi/4$. For even larger cone angles, the CMC signals deviate slightly (about 2-3~\%) due to the influence of higher order terms included in the numerical calculation, which are neglected in the gradient expansion. We can conclude here that the first nontrivial order in the gradient expansion reproduces the numerical calculations exceptionally well, when SOC is only of moderate strength.

The dependence of the CMC signal on the magnitude of $q$ is shown in \Cref{Fig3}~d) for a fixed value of $\Gamma$. Clearly, the spin-spiral pitch  influences the magnitude of CMC in the most drastic way. 
The data reveals that CMC vanishes in the collinear ferromagnetic (FM) limit, $q=0$, and then quickly increases in magnitude over the whole energy range, as $q$ is increased, with the effect being particularly prominent at around $\pm 2$\,eV. Although the qualitative distinction between $\sim q$ and $\sim q^2$ contributions to the longitudinal conductivity of domain walls has been discussed in the past based on model arguments and material-specific calculations~\cite{Kelly2012,FreimuthSeemann2012,MertigDomainWallMR2002,PhysRevB.59.138}, the true chiral nature of the CMC manifests in the dependence of the overall signal on the sense of chirality of the spiral states  observed here.  Remarkably, the chiral magnetoconductivity may change sign several times for different values of $q$ at a fixed value of the Fermi energy. Such complex behavior finds its roots in a strong influence of the spin-spiral vector on the electronic structure of the system, with the corresponding shifts of the bands occurring over the range of eVs.
CMC therefore directly stems from the presence of strong spin-orbit coupling, tying the rearrangements of bands to the sense of chirality as it is visible in \cref{Fig2Z}.
The extreme sensitivity of CMC to wavevector $q$ suggests that for spin-textures hosted in materials with strong spin-orbit interaction, the measured longitudinal conductivity can fluctuate dramatically even upon small variations in  spin distribution, brought about, for example, by an external magnetic field. Naively, such changes in measured magnetotransport can be erroneously interpreted as arising from qualitative modifications in texture properties.

\subsection{Chiral planar Hall effect}

Traditionally, the planar Hall effect is understood as the $\sigma_{(xy)}$ entry of the conductivity tensor, which we find in the second component of irrep $E_2$, $\sigma_{(xy)}=\sigma_{E_{2,2}}[\hatn]$.
For the case of $C_{6v}$ symmetry and the cone spiral as defined before, one finds the chiral contribution as a sum of two terms
\begin{align}
\nonumber \sigma_{E_{2,2}}[\hatn]= &\frac{1}{8}\gPHE \sin ^2(\alpha )(3 \cos(2 \alpha )+1)\sin(2\beta - 2 \phi_q )\\
                                 &- \frac{1}{2} \gCPHE q \sin ^3(\alpha ) \sin(\beta - 2 \phi_q ).
\end{align}
The non-chiral contribution vanishes exactly.
For tilted conical N\'eel spirals, $\beta=0$, the angular dependence is $\sin(2 \phi_q)$ for both terms. As a direct consequence the tensor component describing the CPHE for N\'eel spirals is only nonzero for $\phi_q\neq0$, i.e. when the wavevector $\vec{q}$ is not parallel to the $x$-axis. In contrast, the term in $\sigma_{E_{2,2}}[\hatn]$ proportional to $\gCPHE$ survives for the corresponding spirals of Bloch-type along the $x$-axis, $\beta=\pi/2$ and $\phi_q=0$. This can be understood by realizing that for $\beta=\pi/2$, the term linear in $\gCPHE$ is proportional to $\cos(2 \phi_q)$, while the term linear in $\gPHE$ is proportional to $\sin(2 \phi_q)$. This effective phase shift of $\pi/2$ results in a non-zero contribution to the symmetric transverse conductivity $\sigma_{(xy)}$. Therefore, we investigate $\sigma_{(xy)}$ for the Bloch spiral along the $x$-axis, which has anisotropic contributions  from irrep $E_2$ proportional to $\gCPHE$ only. This is in contrast to the case of $\sigma_{xx}$ calculated for N\'eel spirals along $x$, where both anisotropic and isotropic components from irreps $A_1$ and $E_2$ contribute.

The  band structure of the model for a Bloch spiral with $\pm\mathbf{q}$ along $x$ and the magnitude of $|\vec{q}|=0.31\,a_0^{-1}$, presented in \Cref{Fig2Z}~b), exhibits certain chiral asymmetry of the bands. In comparison to the band structure of the N\'eel spiral shown in \Cref{Fig2Z}~a), the rearrangement of bands is not as drastic for the Bloch spiral. Nonetheless, the chiral PHE is very prominent, as can be observed for a Bloch spiral propagating along the $x$-axis with $|\vec{q}| = 0.47\,a_0^{-1}$ in \cref{Fig3}~b). Similar to the CMC-signal shown in \cref{Fig3}~a), most pronounced features in the CPHE originate from the energy regions around $\pm\SI{1.3}{eV}$ and around $\pm\SI{2.0}{eV}$. While these features are generally broader in energy and similar in magnitude to CMC, strong fingerprints of CPHE appear also around $\SI{0}{eV}$. In comparison to CMC, chiral PHE is not as rapidly decreasing with increasing $\Gamma$, although the overall trend remains.

Moving on to \cref{Fig3}~d), we inspect CPHE as a function of $q$ for a fixed value of $\Gamma$. We observe that the CPHE signal vanishes for the FM limit and then increases in magnitude, but not as rapidly as the CMC signal does. Moreover, at fixed $E_f$,  sign changes with $q$  occur only at very specific energies from the range around $\pm \SI{1}{eV}$ for the CPHE, and not over wide ranges of energies as is the case for CMC. This is a consequence of an overall smoother distribution of CPHE as a function of $E_f$ and $q$. We expect that CMC and CPHE are clearly distinguishable experimentally in materials displaying spiral textures, since they display opposite signs over a wide range of spiral pitch $q$.
One of the most interesting features of the planar Hall effect for Bloch-type spirals propagating along the $x$-axis is suppressed non-chiral PHE, which is given by $\sigma_{(xy)}^{nc}$. This numerical result is in perfect agreement with the predictions from the gradient expansion, see \Cref{tab:TypesHE}. The non-chiral PHE vanishes completely, while we observe a planar Hall effect, comparable in magnitude to AMC, which is purely chiral in nature. 

\subsection{Chiral Hall effect}

We finally discuss chiral corrections to the AHE exhibited by spin-spiral states. 
The AHE is described by the antisymmetric conductivity $\sigma_{[xy]}$, where we find the chiral term
\begin{equation}
    \gCHE  \braket{(\nabla \cdot \hatn ) ( n_\perp^2 - n_\parallel^2 )} 
    =\frac{1}{2} \gCHE  q \cos \beta  \sin (\alpha ) \sin ^2(2 \alpha ) ,
\end{equation}
for the cone spiral as introduced before.
Correspondingly, this effect is referred to as the chiral Hall effect (CHE). 
It is zero for $\alpha =\ang{0}$ and $\alpha =\ang{90}$, and it attains a maximum value at $\alpha \approx \ang{50.77}$. 
We inspect the CHE-signal for the cone-type spiral at the intermediate value $\alpha = \SI{0.6}{rad}\approx\ang{34.38}$. 

In \cref{Fig3}~c) we present the calculated CHE-signal, as given by $\sigma_{[xy]}^{c}$, for the cone-type spiral propagating along the $x$-axis with $|\vec{q}| = 0.47\,a_0^{-1}$ and a cone angle $\alpha = \SI{0.6}{rad}$ as a function of Fermi energy $E_f$ and broadening $\Gamma$. The overall observed structure of the CHE as a function of the Fermi energy is much richer than that exhibited by CMC and CPHE, as well as the non-chiral part of the AHE, $\sigma_{[xy]}^{nc}$ (shown in the inset). Consequently, CHE is characterized by a stronger sensitivity to finer details of the electronic structure mediated by the sense of chirality. In comparison to CMC and CPHE, the most remarkable feature of both chiral and non-chiral parts of the AHE is a qualitatively different behavior with $\Gamma$: in contrast to the data shown in \cref{Fig3}~a)-b) for CMC and CPHE, the CHE-signal does not scale as dramatically with the broadening. Especially the most pronounced features of the CHE located in the energy region $[\SI{-1}{eV},\SI{1}{eV}]$ are least sensitive to $\Gamma$ variation. This behavior can be explained by the robustness of intrinsic contributions to the CHE. In fact, with decreasing disorder the values of the CHE converge to the clean limit values, which are given by the integrated values of Berry curvature $\Omega_{xy}$ antisymmetrized with respect to $\mathbf{q}$~\cite{Kipp2021chiralComPhys}. The rise of non-vanishing intrinsic chiral AHE becomes apparent from looking at \cref{Fig2Z}~c)-d), where the strongly chirality-dependent electronic structure and Berry curvature of the states is visible. On the other hand, by comparing the chiral and non-chiral parts of the AHE, we observe that the disorder corrections to the intrinsic conductivity are much more important for the CHE than for the conventional AHE: the convergence to the clean limit values is quite slow and the saturation does not occur up to broadening $\Gamma$ on the scale of $\SI{0.01}{meV}$. This points to possibly very prominent extrinsic contributions to the chiral Hall effect in realistic spin textures. 

\begin{figure}[t!]
	    \includegraphics[width=0.9\linewidth]{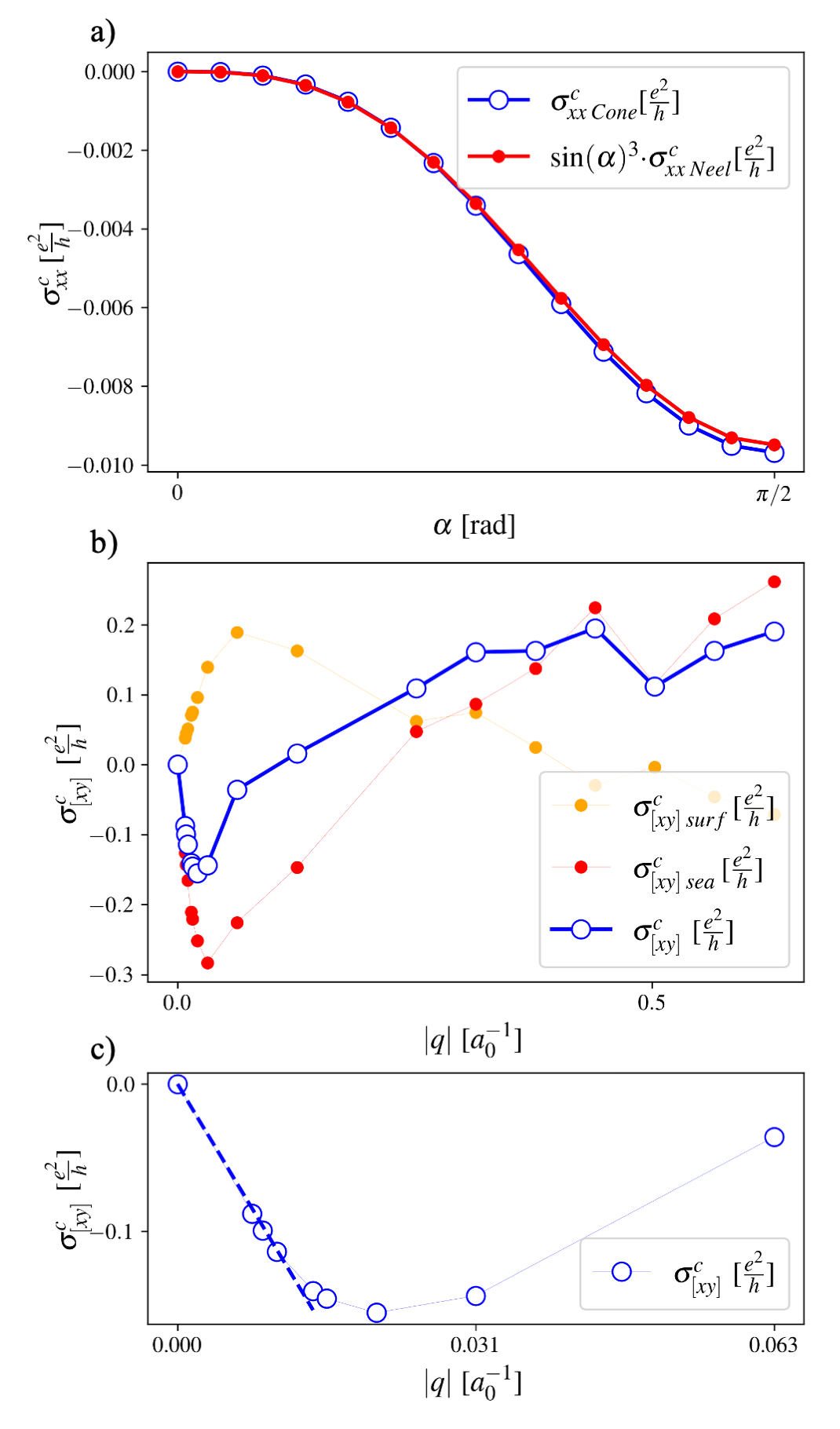}
		\caption{{\bf Scaling with the cone angle and spin-spiral pitch.}
		(a) Chiral  magnetoconductivity (CMC) for a cone-type spiral (blue) calculated for cone angles in the range $\alpha\in[0,\pi/2]$ and CMC for a N\'eel-type spiral (red) calculated for cone angle $\alpha=0$ and extrapolated to larger cone angles by multiplying with $\sin(\alpha)^3$, see \Cref{eq:cos3}. The relationship between CMC for cone-type and N\'eel spirals obtained from the gradient expansion is reproduced with explicit tight-binding calculations. The calculation was done for spin-orbit coupling $\alpha_R = \SI{0.04}{eV}$ at a spiral pitch $q=0.031\,a_0^{-1}$. (b) Ferromagnetic limit of the chiral Hall effect (CHE) for a cone-type spiral. Surface term (orange), sea term (red) and the sum of the two (blue) yielding the CHE for non zero broadening $\Gamma$. 
		(c) The CHE signal near the FM limit of $q=0$.
		The linear slope  $\gamma\approx\SI{-10.752}{e^2a_0/h}$ in the limit of $q\rightarrow 0$, participating in gradient expansion prediction \Cref{eq:linslope_2}, can be estimated from calculations.
		For all plots, the data were calculated for the values of $E_f=\SI{-0.625}{eV}$ and $\Gamma=\SI{0.2}{eV}$. A value of $\alpha_R = \SI{0.4}{eV}$ was used in (b) and (c).
		\label{Fig4}}
\end{figure}

In \cref{Fig3}~f) we analyze the behavior of the CHE signal with the magnitude of the wave-vector $q$.
What can be concluded from comparing all the effects among each other is that while the magnitude of the AHE and CHE is expectedly significantly smaller than that of the CMC and CPHE (at the same value of $\Gamma$), the magnitudes of the chiral and non-chiral AHE are comparable. 
Among the latter two, the CHE is much more sensitive to $q$, which is especially visible for smaller values of $q$, where the structure of the chiral signal is especially non-trivial with respect to the Fermi energy. Here, small deviations from the zero-$q$ limit may result in sign changes in CHE, which bares relevance for magneto-transport detection of textures (see discussion of CMC). 
By inspecting the $q=0$ limit more closely in \cref{Fig4}~b), we find that the data correctly reproduces the ferromagnetic  limit, where the CHE is vanishing. The data presented in \cref{Fig4}~c) also illustrates the linear scaling behavior of the CHE signal $\sigma_{[xy]}^a$ with increasing $q$ away from the ferromagnetic limit. This stands in agreement  with the gradient expansion prediction, and confirms that  higher-order beyond-linear corrections to the chirality-sensitive signal are negligible in the limit of vanishing $q$. 

The limit on the validity of the gradient expansion technique is apparent from the calculations: deviations from linear-in-$q$ behavior visible above $q\approx 0.03\cdot\SI{}{a_0^{-1}}$ $-$ a value which depends significantly on the parameters of the model.
Importantly, in this range of the $q$-vectors we observe that the non-chiral corrections to the $q=0$ values of the AHE are significantly smaller in magnitude than the chiral signal. This bares consequences for our estimates of the qualitative importance of the linear versus higher-order in the magnetization gradients contributions to the transport properties of the large-scale textures, as discussed below.
\section{Impact on transport properties of skyrmions}

While super cell calculations for one-dimensional textures are already computationally demanding, treating two-dimensional textures such as a lattice of skyrmions requires even more resources.
One can however deduce certain transport properties in general slowly varying noncollinear textures from the properties of the spiral phase. 
For example, we consider a skyrmion texture defined by
$\theta(r) = \pi ( 1 - r / w)$, and $\hatn_\mathrm{sk} = (\sin (\theta) \cos (\varphi + \eta), \sin \theta \sin (\varphi + \eta), \cos(\theta))^T $ where $(r, \varphi)$ are polar coordinates in the plane. 
We imagine that the skyrmion $\hatn_\mathrm{sk}$ is embedded in a ferromagnetic host $\hatn_\mathrm{fm} = \vec{e}_z$ and integrate the real-space averages in Eq's. (\ref{eq:local_results_1})-(\ref{eq:local_results_3})
within the volume of the skyrmion $\mathcal{V} = \pi w^2$.
The helicity $\eta$ interpolates between the N\'eel skyrmion for $\eta=0$ and the Bloch skyrmion with $\eta = \pi/2$.
We can then estimate the change in conductivity which is induced by the skyrmion:
\begin{align}
    \sigma_{A_1}^\mathrm{sk}
    - \sigma_{A_1}^\mathrm{fm}
    & = -\frac{\pi}{w} \gCMC \cos \eta
\\
\sigma_{A_2}^\mathrm{sk}
    - \sigma_{A_2}^\mathrm{fm}
    &
    = \left( \frac{4}{\pi^2} - 1 \right) \gAHE - \frac{32}{9\pi w}\gCHE \cos \eta
\\
\sigma_{E_2}^\mathrm{sk}
    - \sigma_{E_2}^\mathrm{fm}    & = 0.
\end{align}
The explicit helicity dependence is one of the hallmarks of ``chiral'' transport effects~\cite{Redies2020}.
In contrast to the topological Hall effect, the CHE is further characterized by its $1/w$ scaling dependence in the long-wavelength limit. 
In the case of the THE, a $1/w^2$-scaling would be expected, originating from the emergent magnetic field $B_\mathrm{eff} \sim \hatn \cdot (\partial_x \hatn \times \partial_y \hatn)$ and which, for large $w$, will eventually be the subdominant contribution of the two different effects.  
Concerning practical computations,
the coefficient $\gAHE$ can be simply extracted from a calculation of a collinear state, while the material parameters $\gCMC$ and $\gCHE$ could be estimated from the spiral calculations as presented before via
\begin{align}
\gCMC &\approx \lim\limits_{q\to 0 } \partial_q
    \frac{  \sigma_{A_1} [\hatn ] 
     }{\cos \beta \sin^3 \alpha  } ,
     \label{eq:linslope_1}
     \\
    \gCHE& \approx \lim\limits_{q\to 0 } \partial_q
    \frac{  \sigma_{A_2} [\hatn ] 
     }{\cos \beta \sin \alpha \sin^2 (2 \alpha)/2 } .
     \label{eq:linslope_2}
\end{align}
The relation is approximate since strong spin-orbit coupling would lead to higher-order corrections in the $\alpha$ and $\beta$ dependence which could not be factored out. 
Irrespective of this, the large effective values of $\gamma$ correspond to the observation that CHE reaches values comparable in magnitude to the collinear AHE already for small values of pitch $q$. This speaks to the fact that the chiral transport signal should be the  dominant source of texture-driven conductivity for wide regimes of texture profiles.

\section{Conclusions}
In this work, we have pursued the idea that homogeneous spin-spiral states can exhibit macroscopic transport properties that are sensitive to their sense of winding, or, chirality. Inspired by previous effective analysis, we referred to an explicit two-dimensional electronic model and rigorously demonstrated the emergence of chiral magnetoconductivity, the chiral planar Hall effect and the chiral Hall effect. Based on Kubo and gradient expansion techniques we showed that the corresponding chiral signal can be very prominent, depending on the type of the spiral state.
It also follows from our analysis that in addition to the high sensitivity of the effects to the electronic structure, the length-scale of the one-dimensional spin texture has a profound effect on the magnitude and sign of the chiral signal. This can prove extremely useful in dealing with aspects such as transport signatures of texture dynamics or magnetic phase transitions. Besides bringing fundamental novel insights into the interplay between spin chirality and magnetotransport, we  show how an ability to predict the underlying properties of simpler spiral states paves the way to understanding and educated engineering of transport fingerprints for more complex textures such as magnetic skyrmions. 

\section*{Acknowledgements}
We  acknowledge  funding  under SPP 2137-"Skyrmionics" of the DFG (German Research Foundation).
We  also gratefully acknowledge the J\"ulich Supercomputing Centre and RWTH Aachen University for providing computational resources under project Nos. jiff40 and jpgi11. 
The work was also supported by the Deutsche Forschungsgemeinschaft (DFG, German Research Foundation) $-$ TRR 173 $-$ 268565370 (project A11), TRR 288 – 422213477 (project B06), and project MO 1731/10-1 of the DFG. J.K. acknowledges funding under HGF-RSF Joint Research Group “TOPOMANN”, Grant No. DB001827. The authors express their gratitude to Prof. Stefan Bl\"ugel and Dr. Juba Bouaziz for fruitful discussion.


\hbadness=99999
\bibliography{literature}

\end{document}

%% file: commands.tex
\newcommand{\gLR}{\gamma_\text{\tiny LC}}
\newcommand{\gMR}{\gamma_\text{\tiny MC}}
\newcommand{\gAMR}{\gamma_\text{\tiny AMC}}
\newcommand{\gCMC}{\gamma_\text{\tiny CMC}}
\newcommand{\gPHE}{\gamma_\text{\tiny PHE}}
\newcommand{\gCPHE}{\gamma_\text{\tiny CPHE}}
\newcommand{\gCHE}{\gamma_\text{\tiny CHE}}
\newcommand{\gAHE}{\gamma_\text{\tiny AHE}}

\newcommand{\pgi}{Peter Gr\"unberg Institut and Institute for Advanced Simulation,
Forschungszentrum J\"ulich and JARA, 52425 J\"ulich, Germany}

\newcommand{\aachen}{Department of Physics, RWTH Aachen University, 52056 Aachen, Germany}

\newcommand{\mainz}{Institute of Physics, Johannes Gutenberg University Mainz, 55099 Mainz, Germany}

\renewcommand{\vec}[1]{\mathbf{#1}} 
\newcommand{\nuv}[1]{\hat{\boldsymbol{#1}}} 
\newcommand{\myMat}[1]{\mathcal{#1}}
\newcommand{\hatn}{\hat{\mathbf{n}}}

\newcommand{\abs}{\mathrm{\abs}}

\newcommand{\dd}{\mathrm{d}}